\documentclass[12pt]{article}

\usepackage[utf8]{inputenc}
\usepackage{roboto}

\usepackage[nodisplayskipstretch]{setspace}
\usepackage{lscape}
\usepackage[bottom=1in,top=1in,left=1in,right=1in]{geometry}
\usepackage{indentfirst}
\usepackage{soul}
\usepackage{titlesec}
\usepackage{appendix}

\usepackage{multirow}

\usepackage{enumerate}
\usepackage[shortlabels]{enumitem}
\setlist[itemize]{topsep=0pt}
\setlist[enumerate]{topsep=0pt}

\usepackage{colortbl}
\usepackage[dvipsnames]{xcolor}
\definecolor{blue}{RGB}{0,0,153}
\definecolor{red}{RGB}{204,0,0}
\definecolor{yellow}{RGB}{255,255,153}
\definecolor{green}{RGB}{0,102,0}
\definecolor{orange}{RGB}{255,128,0}
\definecolor{purple}{RGB}{102,0,204}
\definecolor{orange_graph}{RGB}{245, 121, 58}
\definecolor{blue_graph}{RGB}{15, 32, 128}
\definecolor{MyBackground}{RGB}{255,255,255}
\definecolor{informal}{HTML}{E69F00}
\definecolor{formal}{HTML}{0072B2}
\definecolor{selfemployed}{HTML}{009E73}
\definecolor{unemployed}{HTML}{999999}
\definecolor{zesty1}{HTML}{F5793A}
\definecolor{zesty2}{HTML}{A95AA1}
\definecolor{zesty3}{HTML}{85C0F9}
\definecolor{zesty4}{HTML}{0F2080}

\newcolumntype{a}{>{\columncolor{blue!15}}c}

\usepackage{amssymb,amsfonts,amsthm}
\usepackage[fleqn]{amsmath}
\usepackage{mathtools}
\usepackage{bbm}
\usepackage[multiple]{footmisc}
\usepackage{ulem}

\usepackage{graphicx,graphics,url,caption}
\usepackage{subcaption} 
\usepackage{float}
 
\usepackage{tikz}
\usetikzlibrary{patterns}
\usepackage{array,booktabs,arydshln,xcolor}
\usetikzlibrary{decorations.pathreplacing,angles,quotes,patterns,positioning,shapes.geometric,shapes,snakes,plotmarks}

\usepackage{adjustbox}
\usepackage[flushleft]{threeparttable}

\usepackage[round,authoryear]{natbib}
\setcitestyle{citesep={,}}

\usepackage{hyperref}

\usepackage{siunitx}
\newcolumntype{d}{S[input-symbols = ()]}

\usepackage{multirow}
\newcommand{\PreserveBackslash}[1]{\let\temp=\\#1\let\\=\temp}
\newcolumntype{C}[1]{>{\PreserveBackslash\centering\arraybackslash}m{#1}}
\newcolumntype{R}[1]{>{\PreserveBackslash\raggedleft}m{#1}}
\newcolumntype{L}[1]{>{\PreserveBackslash\raggedright}m{#1}}

\newcommand{\var}{\text{Var}}
\newcommand{\cov}{\text{Cov}}

\usepackage{listings}
\usepackage{color} 
\definecolor{mygreen}{RGB}{28,172,0} 
\definecolor{mylilas}{RGB}{170,55,241}

\usepackage{fp}


\begin{document}

\title{\vspace{-1cm}\textbf{There must be an error here! Experimental evidence on coding errors’ biases}\footnote{We would like to thank Maria Ruth Jones and staff from the World Bank for  valuable insights and cooperation with the project. We are also in debt with Dalila Figueiredo and Eduardo Ferraz for their insightful comments.
The pre-analysis plan for this paper was pre-registered and it is available at the AEA Registry --- AEARCTR-0008312.}}
\author{Bruno Ferman \footnote{Sao Paulo School of Economics - FGV}\and Lucas Finamor\footnote{Sao Paulo School of Economics - FGV}}

\date{\small First draft: August 2025 \\ This draft: September 2025}

\maketitle

\begin{abstract}

Quantitative research relies heavily on coding, and coding errors are relatively common even in published research.  In this paper, we examine whether individuals are more or less likely to check their code depending on the results they obtain. We test this hypothesis in a randomized experiment embedded in the recruitment process for research positions at a large international economic organization. In a coding task designed to assess candidates’ programming abilities, we randomize whether participants obtain an expected or unexpected result if they commit a simple coding error. We find that individuals are almost 20\% more likely to detect coding errors when they lead to unexpected results. This asymmetry in error detection depending on the results they generate suggests that coding errors may lead to biased findings in scientific research.
\end{abstract}


\onehalfspacing

\section{Introduction}

There is growing concern about the lack of reproducibility in scientific research, both in economics and other fields
(\cite{
ankel2023economists,
brodeur2024,
brodeur2024promoting,
brodeur2025assessing,
camerer2016evaluating,
camerer2019replication,
campbell2024,
chang2022,
christensen2018,
dewald1986,
drazen2021journal,
gertler2018make,
hamermesh2007replication,
lang2025,
mccullough2006lessons,
osc2015,
perignon2024computational,
vilhuber2019report,
wood2018push}).
Several factors contribute to this problem, including the unavailability or poor documentation of data and code (\cite{
anderson2008replication,
chang2022,
dewald1986,
gertler2018make,
mccullough2007got,
vilhuber2019report}), p-hacking and publication bias  (\cite{
adda2020p,
andrews2019identification,
ashenfelter1999review,
AshenfelterGreenstone2004,
begg1994operating,
blancoperez2020publication,
brodeur2016star,
brodeur_methods,
brodeur2022mturk,
Brodeur_unpacking,
brodeur2024job,
brodeur2024phacking,
brodeur2024preregistration,
bruns2019reporting,
camerer2016evaluating,
campbell2024,
CardKrueger1995,
chopra2024null,
dellavigna2022rcts,
delong1992all,
doucouliagos2013all,
dreber2024selective,
elliott2022detecting,
franco2014publication,
gerber2008publication,
gerber2008statistical,
gerber2010publication,
havranek2015measuring,
henry2009strategic,
ioannidis2005most,
ioannidis2017power,
kepes2022questionable,
leamer1983reporting,
mccloskey1985loss,
oboyle2017chrysalis,
olsen2019research,
rosenthal1979file,
stanley2005beyond,
stanley2008meta,
vivalt2019specification}), excessive researcher degrees of freedom in data construction and analysis (\cite{
huntington2021influence,
huntington2025,
landy2020,
menkveld2024nonstandard,
silberzahn2018many,
simmons2011false}), and coding errors that can alter empirical conclusions (\cite{
anderson2008replication,
brodeur2024,
mccullough2006lessons,
mccullough2007got}). In some cases, coding errors have led to widely cited but incorrect findings—for example, the paper by \cite{reinhart2010growth} on debt and growth, which omitted data due to a spreadsheet error, or replications of published articles that uncovered computational mistakes affecting key results (\cite{herndon2014does}). Beyond these anecdotal examples, in a mass replication study, \cite{brodeur2024} find that one quarter of studies published after 2022 in nine leading economics journals and three leading political science journals had some coding errors, showing that coding errors are highly prevalent.

If coding errors are independent of the results they generate, they would not lead to systematic bias, although they would still contribute to the excess dispersion of estimates observed in empirical research. In this case, it would just be part of what some have called “nonstandard errors” (\cite{
huntington2021influence,
huntington2025,
menkveld2024nonstandard,
silberzahn2018many}). This variation goes beyond what is typically captured by standard model- or design-based measures of uncertainty and may arise from researcher practices and flexibility in analytical choices, where the possibility of coding errors would be one of the reasons why different groups of researchers may end up with different results. In contrast, if the likelihood of detecting coding errors depends on the results those errors produce, then, in addition to increasing dispersion, even well-intentioned researchers may unknowingly introduce systematic bias into their estimates due to coding errors.

In this paper, we test the hypothesis that the probability of detecting a coding error depends on the outcome the error generates. We do so through a randomized experiment embedded in the recruitment process for research positions at a large international economic organization. As part of a coding task designed to assess candidates’ programming abilities, we randomize whether a simple coding mistake leads to an expected or unexpected result. The coding mistake is whether individuals take into account that the value 99 in the outcome variable codes missing values. Failing to take into account the missing values leads to wrong results that can be expected or unexpected, depending on the group candidates were randomly allocated. This design allows us to estimate whether individuals are less likely to detect and correct coding errors when the resulting outcome aligns with their expectations. We find that individuals are almost 20\% more likely to detect coding errors when they generate unexpected results. This asymmetry in error detection suggests that coding mistakes may lead not only to an increase in dispersion that is not captured by usual standard errors but also to bias in empirical research.

Our findings contribute to several strands of the literature. First, we add to the growing body of work on the reproducibility crisis in economics by highlighting a novel behavioral mechanism—selective error detection—that can undermine the reliability of empirical findings even in the absence of intentional misconduct. Second, we speak to the literature on nonstandard errors showing that undetected coding errors may be an underappreciated source of excess variation in empirical estimates, showing it can affect not only dispersion but also introduce bias on estimators. Finally, our results echo themes from the literature on confirmation bias (\cite{kunda1990, nickerson1998confirmation}), suggesting that researchers may be more likely to overlook errors that produce results aligned with prior expectations.

\section{Experimental Design}

\subsection{Setting and sample}

The study takes place in a recruitment process for research assistants and for a research-oriented fellowship program within the Development Economics (DEC) Vice Presidency of the World Bank in two separate waves in 2024 and 2025.
As part of the recruitment process, candidates are asked to perform a simple data task to evaluate their coding abilities. The experiment takes place within this data task.

The data task was the last component of the first screening performed by the recruiters. Completion of the task was encouraged, but it was not a requirement.\footnote{Some positions did not require coding, therefore the coding task was not mandatory.} At the start of the test, individuals could decide whether to share their data from the test for research purposes. The decision had no impact on how the data task was used for the recruitment process. Pooling the two waves, we have results for 1,036 task takers who started the data task and agreed to share their data for research purposes.\footnote{ In total, 1,171 task takers started the data task and answered the initial questions. Among them, 135 (11.5\%) opted to not share their data.}

\subsection{Data task}

The data task presented candidates with a scenario in which they had to analyze data from a hypothetical RCT intervention that tailored educational content to students’ appropriate level---an intervention inspired by studies such as \cite{banerjee2007,cabezas2011,duflo2011} and \cite{banerjee2016}. The main objective was to manipulate whether a coding error would lead to an expected or unexpected result, allowing us to evaluate whether participants are more likely to debug their code when they observe an unexpected outcome. To do so, we create  datasets in which missing values for the outcome variable (test scores) were coded as 99---an information disclosed in the data dictionary. We then experimentally varied whether including students with missing outcomes in the regressions would produce expected or unexpected results. We describe the data task in detail below, highlighting the features relevant to our research design. In Appendix~\ref{Appendix_datatask}, we reproduce the data task.

The data task began with a set of initial demographic questions, including gender, education level, whether the candidate had taken an econometrics course, and the language in which they intended to complete the task. These variables were used as individual-level controls and to conduct heterogeneity analysis.

Next, candidates were presented with results from six randomized controlled trials that evaluated the impact of programs tailoring educational content to students’ appropriate level.\footnote{The results were drawn from the following papers: \cite{banerjee2007,cabezas2011,duflo2011}, and \cite{banerjee2016}. {All candidates observe the results from all articles.}} The estimated effects ranged from 0.08 to 0.16 standard deviations---each positive and statistically significant at the 5\% level. These results served to anchor participants’ beliefs, reinforcing the expectation of positive effects from interventions of this type.

We then presented a hypothetical RCT of an intervention that, inspired by the literature, tailored educational content to students’ appropriate level. We explained that 5th-grade teachers in treated schools received materials and training to implement the tailored program, while instruction in control schools remained unchanged. At this stage, participants were asked to report their best guess of the approximate effect of such an RCT on language proficiency, measured 12 months after the start of the intervention. Using a slider, they could select values between –0.30 and 0.30 standard deviations. Consistent with the anchoring provided by existing evidence, more than 87\% of task-takers reported priors between 0.08 and 0.16 standard deviations.

After collecting participants’ priors, each candidate received a dataset corresponding to the hypothetical experiment. Each task taker was randomly assigned a version of the dataset, which included three files: student-level data from 480 schools participating in the RCT across two distinct states (one file per state), and a data dictionary. Candidates were informed that they would answer four questions based on these datasets. Each question appeared on a separate page, and once they proceeded to the next question, they could not return to change previous answers. However, they were told in advance that they would have the opportunity to review and revise their responses after completing all questions. While only the final answers were used for the screening process, we rely on the initial responses for the purposes of our experiment.\footnote{As we explain in detail in Section \ref{sec:fairness}, this design ensures the experiment is fair to participants both ex-ante and ex-post.}  The questions required either a numerical answer (e.g., a point estimate with three decimal places) or a written response involving interpretation or reasoning.

The first question (Q1) requires task takers to manipulate the data and answer questions related to counts, means, and conditional means based on the provided dataset. The second question (Q2) asks participants to run an OLS regression to assess the balance of the hypothetical RCT. Importantly, the variables used in these two questions do not contain missing values, as the outcome variable (test scores), which contains the missing values coded as 99, is not used. We refer to the scores obtained in these initial questions (Q1--Q2) as a measure of initial coding ability, as they evaluate whether individuals can perform basic data manipulations and run a standard OLS regression. As specified in our pre-analysis plan, our main analyses are restricted to individuals who demonstrate basic coding proficiency.\footnote{The main reason for this restriction is that we depend on individuals knowing how to run an OLS regression in order to correctly identify whether they have spotted the coding error.}

The third question (Q3) asks participants to estimate the effect of the hypothetical intervention on language scores in one of the two states (e.g., State 1). Task takers are instructed to run a specific OLS regression, using the standardized language score as the outcome and regressing it on a constant and a treatment indicator. We record their submitted point estimate, standard error, and p-value for the treatment effect, as well as their interpretation of the result. The fourth question (Q4) mirrors this task for the other state (e.g., State 2). That is, if a participant answered Q3 using data from State 1, Q4 asks about State 2, and vice versa. In these two questions, participants are exposed to a common coding error: failing to account for the fact that missing test scores are recorded as 99. The data dictionary explicitly informed them that a value of 99 corresponds to missing outcomes.

The key experimental variation lies in the construction of the datasets provided to candidates: they differ in the results participants would obtain if they do \textbf{not} account for the coding of missing values. In the treatment group, failing to drop the 99s leads to a significant negative estimated effect of the program in Q3 (an unexpected result given their priors), followed by a significant positive effect in Q4 (an expected result). In contrast, the control group encounters the reverse: a significant positive result in Q3 and a significant negative result in Q4, if the 99 values are mistakenly included in the regressions. If the missing values are appropriately taken into account, then the estimated effects are approximately zero in both groups. Table \ref{tab:experimental_design} shows the average point estimate candidates would obtain in both questions if they include the 99 values or not.

\begin{table}[h]
        \centering
\begin{threeparttable}
    \caption{Experimental Design}\label{tab:experimental_design}
    \begin{tabular}{lcc}
    \hline \hline
         & Treated & Control  \\ \hline 
        \addlinespace
        \multicolumn{3}{l}{\textbf{Panel A. Keeping missing values (99)}} \\ 
        \addlinespace
         Question 3 & $-0.1602^{***}$   &  $0.1488^{***}$ \\
         Question 4 & $0.1488^{***}$    & $-0.1602^{***}$ \\ 
         \\
         \multicolumn{3}{l}{\textbf{Panel B. Removing missing values (99)}} \\ 
         \addlinespace
         Question 3 &    -0.0042  &  0.0068  \\
         Question 4 &  0.0068 & -0.0042   \\ 
         \hline \hline 
    \end{tabular}
    \begin{tablenotes}
        \item \footnotesize{Notes: The table shows the average value of the expected coefficient test takers would obtain in questions 3 and 4, depending on whether they were randomized into the treatment or control group, and whether they have removed the missing values (coded as 99).}
    \end{tablenotes}
    \end{threeparttable}
\end{table}

On the final page of the data task, candidates are shown all their previous answers and are given the opportunity to revise them before submission. For instance, a candidate who notices the issue with the 99s only in Q4 could still revise their answer to Q3 before submitting. Importantly, for the purposes of our experiment, we analyze the sequential responses prior to these final adjustments. In contrast, the screening process considers only the final submitted answers.

\subsection{Identification}\label{sec:identification}

The main goal of the experiment is to identify the proportion of individuals who only spot the coding error when this leads to an unexpected result. Our experiment design, with  variation on whether the unexpected result (in case the coding error is not spotted) appears in Q3 or Q4 allows us to identify this proportion in two different ways. 

We classify the individuals according to four latent types:

\begin{enumerate}
    \item Always-spot (AS): those who always spot the error, irrespectively of the result;

    \item Never-spot (NS): those who never spot the error, irrespectively of the result; 

    \item Complier I (CI): those who spot the coding error if it leads to unexpected results;

    \item Complier II (CII): those who spot the coding error if they find conflicting results between the two answers (that is, Q3 had a positive effect while Q4 had a negative one, or vice-versa).
    
\end{enumerate}
Figure \ref{fig:latent_types} presents the observed outcomes for questions Q3 and Q4 for each of these latent types, depending on whether they are in the treated or in the control group.\footnote{This classification would not allow individuals to detect the problem in Q3, but not detect it in Q4. Reassuringly, we find that only 2 out of 1,036 subjects presented this pattern.} Given that, we have two ways of identifying the proportion of CI types.

\begin{figure}[!h]
\caption{Latent types and identification}\label{fig:latent_types}
\begin{center}
\begin{minipage}{0.48\textwidth}
\centering
\begin{tikzpicture}
\draw (0,0) rectangle (6,4);
\draw (3,0) -- (3,4); 
\draw (0,2) -- (6,2); 

\node at (1.5, 4.3) {\footnotesize Spotted};
\node at (4.5, 4.3) {\footnotesize Not Spotted};
\node at (3, 4.8) {Q4};
\node at (3, 5.6) {\bfseries Control};

\node[rotate=90] at (-0.3, 3) {\footnotesize Spotted};
\node[rotate=90] at (-0.3, 1) {\footnotesize Not Spotted};
\node[rotate=90] at (-1, 2) {Q3};

\node at (1.5, 3) {AS};
\node at (1.5, 1) {CI, CII};
\node at (4.5, 1) {NS};
\node at (4.5, 3) {-};
\end{tikzpicture}
\end{minipage}
\hfill
\begin{minipage}{0.48\textwidth}
\centering
\begin{tikzpicture}
\draw (0,0) rectangle (6,4);
\draw (3,0) -- (3,4); 
\draw (0,2) -- (6,2); 

\node at (1.5, 4.3) {\footnotesize Spotted};
\node at (4.5, 4.3) {\footnotesize Not Spotted};
\node at (3, 4.8) {Q4};
\node at (3, 5.6) {\bfseries Treated};

\node[rotate=90] at (-0.3, 3) {\footnotesize Spotted};
\node[rotate=90] at (-0.3, 1) {\footnotesize Not Spotted};
\node[rotate=90] at (-1, 2) {Q3};

\node at (1.5, 3) {AS, CI};
\node at (1.5, 1) {CII};
\node at (4.5, 1) {NS};
\node at (4.5, 3) {-};
\end{tikzpicture}
\end{minipage}
\end{center}
\end{figure}

The first approach uses only data from Q3. In the control group, only the AS type would spot the error at this stage, while in the treated both the AS and the CI types would spot the error. Therefore, we can identify the proportion of CI types by comparing the proportion of participants who answered correctly Q3 between the treated and the control groups. 

Another alternative is to consider participants who did not spot the error in Q3, but did spot the error in Q4. As presented in Figure \ref{fig:latent_types}, in the treated group, only type CII would present this observed pattern, while in the control group this pattern would be observed for both CI and CII types. Therefore, a comparison of the proportion of participants that exhibit this observed pattern between the control and the treated groups would provide another way of identifying the proportion of individuals who debug when they observe an unexpected result.\footnote{Here we assume that the order individuals obtain the wrong results, including the 99s, does not matter. In Appendix~\ref{app:latent} we expand the analysis for the cases where this might not hold.} While we did not pre-specify the use of this variation for identifying the proportion of CI types, we observed afterwards that this provides a clean and more precise identification of the parameter of interest.  

\subsection{Fairness and ethical concerns}\label{sec:fairness}

In addition to helping with the identification, the use of questions Q3 and Q4 plays a crucial role in guaranteeing that the experiment is not only \textit{ex-ante} fair for all candidates---which is accomplished given that all candidates had the same probability of being assigned to the treatment or the control group), but also that the experiment is \textit{ex-post} fair---that is, we wanted to place candidates in analogous situations, irrespectively of the results of the randomization. Suppose we had only one question. The main concern was that, if CI types are prevalent in the pool of candidates, then those assigned to the treatment group would have an advantage, as they would be more likely to answer this question correctly. 

With questions Q3 and Q4, we still have the issue that CI types assigned to the treatment group would  realize the 99 issue in Q3---so they would be able to answer correctly Q3 and Q4, whereas those assigned to the control group would only answer correctly Q4. However, candidates are able to edit their answers before submitting them, which eliminates this problem. More specifically, CI types assigned to the control group would only realize that 99 values represent missing values when answering Q4. And then they would be able to edit their answer to Q3 before submitting it for the screening process. The fact that only final answers were used for the screening process, while we rely on initial responses for the experiment, allows the experiment to be \textit{ex-post} fair while still providing the relevant information for identification of the proportion of participants who only spot the error when it leads to an unexpected result.

We also piloted the experiment in four different recruitment processes with a different partner and a similar design. Results did not show any statistical difference in final recruitment scores between treated and control individuals. The pre-analysis plan presents the results for these pilots. In the experiment we also obtain the same result that there is no statistically significant difference in the final score across the two groups. It is also worth mentioning that the recruitment process already measured coding proficiency with randomized questions. The only manipulation for this RCT was on the structure of the dataset and questions.

\subsection{Descriptive Statistics}

Our total sample comprises 1,036 task takers. Among them, 633 (61.1\%) completed the data task in the 2024 wave, and 406 (39.9\%) in the 2025 wave. Every time a candidate started the task, they would be randomized into the control or treatment group. A total of 527 (50.9\%) task takers were randomized into the control group, and 509 (49.1\%) were randomized into treatment. In order to correctly identify whether individuals saw the 99 as coding missing values or not, we depend on individuals knowing how to run a regression. Therefore, we define a subset of our sample, the qualified sample, which correctly estimated an OLS regression in question 2, the question preceding the two questions we use in the experiment. A total of 807 (77.9\%) candidates are in the qualified sample. Our pre-analysis plan specified at least 800 observations in the qualified sample.

Table~\ref{tab:desc_stat} shows descriptive statistics of the sample. The first three columns show the {overall} mean value, mean for the control group, and mean for the treated group for each variable. The next two columns show the estimated difference and p-value for the regression-adjusted balance test of equality of means. The last column shows the number of observations with non-missing information. We can see that 37.7\% of the sample are female, 93.7\% have a master's degree or are enrolled in a master's program, and 86.4\% have already taken an Econometrics course. In terms of computational language, 50.8\% use Stata, 29.5\% R, and 15.9\% Python, and the remaining 3.8\% use other software. The average score in the initial two coding questions is 4.44 out of 6 possible points. In the recruitment process of the partner institute, they also recorded coding proficiency and scored individuals for their knowledge on impact evaluation. Coding score reflects answers to multiple choice questions on the language of their choice (options were Stata, R, or Python). We assigned each individual the corresponding score from the language they choose to complete our data task. We standardized this variable separately for each software language to have mean zero and unitary standard deviation in the control group. Impact evaluation questions measured individuals familiarity and knowledge on impact evaluation and econometric questions. We also standardized this variable to have mean zero and unitary standard deviation in the control group. In terms of beliefs about the effects of the hypothetical RCT, the average value is 0.126 standard deviations, in the middle of the interval of the presented papers. Figure~\ref{fig:prior} shows the histogram for these priors. Consistent with the randomization protocol, we do not find significant differences between treated and control groups in terms of these covariates. The p-value of a joint test that means of all these variables are the same between these two groups is 0.463. {Table~\ref{tab:balance_qualified} reproduces these results for the qualified sample.\footnote{The p-value of a joint test that means of all covariates are the same between treated and control groups is 0.606 for the qualified sample. We do find, however, statistically significant differences in the proportions of individuals with master's degree or higher. As pre-specified in the PAP, we include this (and the other covariates) as controls in our main specifications. Reassuringly, all results remain similar upon the inclusion of covariates. Additionally, Tables~\ref{tab:heterogeneities} and \ref{tab:heterogeneities_app} show results conditioning on these variables.}

\begin{table}[!ht]
    \centering
    \begin{threeparttable}
        \caption{Descriptive statistics and balance}\label{tab:desc_stat}
    \label{tab:my_label}
    \begin{tabular}{lC{1.5cm}C{1.5cm}C{1.5cm}C{1.5cm}C{1.5cm}C{1.5cm}}
    \hline \hline 
    \addlinespace
Variable & Overall Mean & Control Mean & Treated Mean & Diff & P-value Diff & N Obs\\
\addlinespace
\hline 
\addlinespace
Female & 0.377 & 0.352 & 0.402 & 0.051 & 0.095 & 1025\\
 & (0.485) & (0.478) & (0.491) & {}[0.030] &  & \\
 \addlinespace
Master & 0.937 & 0.927 & 0.947 & 0.019 & 0.203 & 1031\\
 & (0.243) & (0.260) & (0.225) & {}[0.015] &  & \\
 \addlinespace
Econometrics & 0.864 & 0.859 & 0.868 & 0.009 & 0.663 & 1034\\
 & (0.343) & (0.348) & (0.338) & {}[0.021] &  & \\
 \addlinespace
Stata & 0.508 & 0.524 & 0.491 & -0.033 & 0.295 & 1036\\
 & (0.500) & (0.500) & (0.500) & {}[0.031] &  & \\
 \addlinespace
R & 0.295 & 0.287 & 0.305 & 0.018 & 0.526 & 1036\\
 & (0.456) & (0.453) & (0.461) & {}[0.028] &  & \\
\addlinespace
Python & 0.159 & 0.150 & 0.169 & 0.019 & 0.403 & 1036\\
 & (0.366) & (0.357) & (0.375) & {}[0.023] &  & \\
 \addlinespace
 Score Q1-Q2 & 4.446 & 4.438 & 4.454 & 0.016 & 0.891 & 991\\
 & (1.828) & (1.850) & (1.807) & {}[0.116] &  & \\
 \addlinespace
Coding Score & 0.055 & 0.000 & 0.110 & 0.110 & 0.109 & 886\\
 & (1.025) & (0.998) & (1.049) & {}[0.069] &  & \\
 \addlinespace
Impact Evaluation Score & 0.032 & 0.000 & 0.064 & 0.064 & 0.328 & 949\\
 & (1.009) & (1.000) & (1.018) & {}[0.066] &  & \\
 \addlinespace
Prior effect & 0.126 & 0.127 & 0.125 & -0.002 & 0.466 & 947\\
 & (0.046) & (0.046) & (0.047) & {}[0.003] &  & \\
\\
P-value joint test & & & & & 0.463 & \\
\addlinespace
\hline \hline 
\end{tabular}
    \begin{tablenotes}
        \item \footnotesize{Notes: The table shows the overall average, the average for the control group, the average for the treated group, the difference between the treated and control groups, the regression-adjusted p-value testing the equality of means for the treated and control groups, and the number of observations with non-missing values for each variable. The number in parenthesis is the standard deviation for each variable. The number in brackets shows the standard error of the regression-adjusted difference. The last row shows the p-value for the joint test that means of all these variables are the same between treated and control groups.  Number of observations for each variable varies due to non-response to the specific question. Non-response rates are balanced between treated and control groups.}
    \end{tablenotes}
        \end{threeparttable}
\end{table}

\section{Empirical Strategy}

Our empirical strategy explores directly the random assignment of the ordering of the negative--positive results on questions 3 and 4. Let $Y^{Q3}_i$ be the indicator for whether the candidate $i$ spotted the error in the first question to estimate the causal effects (question 3). Candidate $i$ is treated, $T_i=1$, if she receives the negative estimate in the first question. We estimate our treatment effects in a specification that interacts the treatment indicator with the demeaned control variables and wave indicators, as suggested by \cite{Lin2013Agnostic},
\begin{equation}
\begin{aligned}
Y^{Q3}_i &= \alpha_1 + \beta_1 T_i  + \gamma_1 \tilde{X}_i + \gamma_2\tilde{W}_i +\gamma_3 T_i \tilde{X}_i + \gamma_4 T_i\tilde{W}_i + \gamma_5 \tilde{X}_i\tilde{W}_i + \gamma_6 T_i\tilde{X}_i\tilde{W}_i   +\varepsilon_i \\
&= \alpha_1 + \beta_1 T_i + \gamma Z + \varepsilon_i ,
\label{main_eq}
\end{aligned}
\end{equation} 
where $\tilde{X}_i$ are all demeaned covariates: gender, whether the candidate took an econometrics course, whether the candidate has a master's degree or above, and the initial score in the screening questions. In addition to the above covariates, we include a control for (demeaned) wave indicator ($\tilde{W}_i$) and the interaction of all covariates and the wave indicator, to account for the potential different sets of candidates in each wave. The second line uses the variable $Z$ for a short notation of all these variables. $\alpha_1$ measures the proportion of individuals in the control group who spot the error. $\beta_1$ is our coefficient of interest, measuring the differential probability of detecting the coding error for individuals observing the negative effect in the first question. We will estimate Equation \ref{main_eq} using OLS with robust standard errors. This is the equation and estimation method pre-specified in our pre-analysis plan. In the pre-analysis plan, we also specified that we would conduct inference using a unilateral hypothesis test. For completeness, we present p-values for both unilateral and bilateral tests.

As discussed in Section~\ref{sec:identification}, it is also possible to identify the proportion of individuals who only spot the coding error when this leads to an unexpected result by comparing the proportion of individuals  who only spot the error in question 4 in the control and treated groups (importantly, in this case it is the proportion of controls minus the proportion of treated). In practice, we can implement this identification strategy using the following regression:
\begin{equation}
\begin{aligned}
\tilde Y^{Q4}_i &= \alpha_2 + \beta_2(1-T_i) + \delta Z + \epsilon_i,
\label{sec_eq}
\end{aligned}
\end{equation} 
where we define $\tilde Y^{Q4}_i$ as one if individual $i$ spotted the error for the first time in Q4 (thus, not in Q3). The coefficient $\alpha_2$ measures the proportion of individuals in the treated sample who spot the error because they saw the flipped results, that is CII, while $\beta_2$ is still the same coefficient of interest in measuring the proportion of individuals who spot the error only because they see the negative result that is the CI type. We also estimate Equation \ref{sec_eq} with OLS with robust standard errors.  

As both $\hat{\beta_1}$ and $\hat{\beta_2}$ are different estimators of the same parameter of interest, we can combine their estimations to achieve a more efficient estimator. We do it in two ways. The first estimator combines both estimates, choosing the weights that minimize the variance. That is,
\begin{equation}
\begin{aligned}
\hat{\beta}_{\text{combined}} = \omega \hat{\beta_1}+(1-\omega)\hat{\beta_2}.
\end{aligned}
\end{equation} 
Where $\omega$ is chosen to minimize the variance of $\hat{\beta}_{\text{combined}}$. That is 
\begin{equation}
\begin{aligned}
\omega=\frac{\var(\hat{\beta_1})-\cov(\hat{\beta_1},\hat{\beta_2})}{\var(\hat{\beta_1})+\var(\hat{\beta_1})-2\cov(\hat{\beta_1},\hat{\beta_2})} \;.
\end{aligned}
\end{equation} 
To take into account that $\omega$ uses estimated variances and covariances, we use bootstrap at the individual level to conduct inference for $\hat{\beta}_{\text{combined}}$. At every bootstrap iteration, we compute $\hat{\beta_1}$ and $\hat{\beta_2}$, their variance-covariance matrix, and then the combined estimator. The resulting bootstrap p-values are very close to the analytical ones obtained ignoring how $\omega$ uses estimated variances and covariances.

Another approach to obtain a more efficient estimator is to jointly estimate Equations \ref{main_eq} and \ref{sec_eq}, imposing the same coefficient $\beta$. We do it using a GMM estimator that stacks the moments of the two equations together. As in the OLS estimator, the moments are that the covariances of residuals and all variables are zero. Estimators $\hat{\beta_2}$,  $\hat{\beta}_{\text{combined}}$ and $\hat \beta_{\text{GMM}}$ were not in our pre-analysis plan as we have not planned to use the data from Q4. After the implementation, we saw how it provides as useful and clean variation as the first question, with a very minimal additional assumption.

\section{Results}

\subsection{Main Result}

Table~\ref{tab:main} presents the main results. In the first column, we present the estimates using the first estimator (presented in Equation \ref{main_eq}). In the first panel, we only add the wave fixed effects. 
We can see that only 7.8\% of the control group identifies the 99 in Q3 (the intercept from  Equation \ref{main_eq}). For the treatment group, the proportion spotting the error is 1.03pp higher, increasing it to 8.8\%, although this estimate is not statistically significant. The next three panels add, sequentially, all the control variables: all demographic controls (Panel B), data test measures (computational language used, and initial proficiency scores) in Panel C, and screening variables (coding and impact evaluation scores) in Panel D. The results are all very similar; the point estimate ranges between 0.79pp and 1.11pp, with relatively large standard errors. 

\begin{table}[!h]
    \centering
    \begin{threeparttable}
        \caption{Main results}\label{tab:main}
    \begin{tabular}{L{4cm}C{2.6cm}C{2.6cm}C{2.6cm}C{2.6cm}C{2.6cm}}
    \hline \hline 
    \addlinespace
Estimator & $\hat{\beta}_1$& $\hat{\beta}_2$& $\hat{\beta}_{\text{combined}}$& $\hat{\beta}_{\text{GMM}}$\\
& (1) & (2) & (3) & (4) \\ 
\addlinespace
\hline 
\addlinespace
\multicolumn{5}{l}{\textbf{Panel A - Controls for wave}}\\
\addlinespace
Treat        & \num{0.0103}       & \num{0.0146}       & \num{0.0141}       & \num{0.0141}       \\
(s.e.) & (\num{0.0199})     & (\num{0.0077})     & (\num{0.0073})     & (\num{0.0073})     \\
\addlinespace

[p-value]& [\num{0.6053}]     & [\num{0.0596}]     & [\num{0.0514}]     & [\num{0.0546}]     \\
\{p-value2\}& \{\num{0.3026}\} & \{\num{0.0298}\} & \{\num{0.0278}\} & \{\num{0.0273}\} \\

\\
\multicolumn{5}{l}{\textbf{Panel B - Adding demographics controls}}\\
\addlinespace
Treat        & \num{0.0100}       & \num{0.0157}       & \num{0.0150}       & \num{0.0149}       \\
(s.e.) & (\num{0.0197})     & (\num{0.0080})     & (\num{0.0075})     & (\num{0.0077})     \\
\addlinespace

[p-value]& [\num{0.6126}]     & [\num{0.0508}]     & [\num{0.0400}]     & [\num{0.0525}]     \\
\{p-value2\}& \{\num{0.3063}\} & \{\num{0.0254}\} & \{\num{0.0213}\} & \{\num{0.0262}\} \\

\\
\multicolumn{5}{l}{\textbf{Panel C - Adding data test variables}}\\
\addlinespace
Treat        & \num{0.0111}       & \num{0.0145}       & \num{0.0141}       & \num{0.0141}       \\
(s.e.)& (\num{0.0198})     & (\num{0.0079})     & (\num{0.0075})     & (\num{0.0077})     \\
\addlinespace

[p-value]& [\num{0.5738}]     & [\num{0.0674}]     & [\num{0.0564}]     & [\num{0.0664}]     \\
\{p-value2\}& \{\num{0.2869}\} & \{\num{0.0337}\} & \{\num{0.0308}\} & \{\num{0.0332}\} \\
\\
\multicolumn{5}{l}{\textbf{Panel D - Adding screening test variables}}\\
\addlinespace
Treat        & \num{0.0079}       & \num{0.0130}       & \num{0.0124}       & \num{0.0124}       \\
(s.e.)& (\num{0.0198})     & (\num{0.0076})     & (\num{0.0072})     & (\num{0.0074})     \\
\addlinespace

[p-value]& [\num{0.6878}]     & [\num{0.0864}]     & [\num{0.0792}]     & [\num{0.0949}]     \\
\{p-value2\}& \{\num{0.3439}\} & \{\num{0.0432}\} & \{\num{0.0380}\} & \{\num{0.0475}\} \\
\\
N Obs        & \num{788}          & \num{788}          & \num{788}          & \num{788}          \\
Intercept & \num{ 0.078} & \num{ 0.005}  & - & -  \\
\addlinespace
\hline \hline 
\end{tabular}
    \begin{tablenotes}
        \item \footnotesize{Notes: The table presents the estimates for the four main estimators, each in one column. The four panels differ according to the control variables included in the estimation procedure: only wave indicators, adding demographics controls, adding data test variables (the language chosen and Score Q1-Q2), and adding screening test variables (coding and impact evaluation scores). Robust standard errors in parentheses. In brackets and curly brackets are the p-values for the bilateral and unilateral tests. For the combined estimator, p-values were obtained using bootstrap with 30,000 replications. The intercept in Column 1 provides an estimator for the proportion of candidates who always spot the coding errors, regardless of the results they generate. The intercept in Column 2 provides an estimator for the proportion of candidates that spot the errors if they produce conflicting results.}
    \end{tablenotes}
        \end{threeparttable}
\end{table}

In the second column, we implement the OLS estimator using the data from question 4. The intercept from this regression reveals that only 0.5\% of individuals in the treated group spot the error only in this question (that is, those who detect the error when they see the flipped result). In the control group, the proportion that only spots the error in Q4 is higher, yielding a $\hat \beta_2$ equal to 1.46pp. It is remarkable how close this estimate is to the estimate using only the first question (1.03pp). Moreover, this estimator is considerably more precise than the previous one. This happens because the proportion of treated individuals who only spot in the second question is very close to zero. Even without additional controls, this estimate is marginally significant with a p-value of 0.0596 for a bilateral test (and statistically significant at 5\% when we consider a unilateral test, with a p-value of 0.0298). When we add controls in the next panels, the estimate is very stable, ranging from 1.30pp to 1.57pp, with p-values for the bilateral test in the interval 5.1\%--8.6\%.

Results in the third column combine optimally the two estimators, given their variances and covariances. The estimate without additional controls is 1.41pp. This is an 18.1\% increase over the baseline detection probability. Including all controls does not change the result. The estimates are between 1.24pp and 1.50pp with p-values ranging between 4.0\% and 7.9\% for the bilateral test and  between 2.1\%--3.8\% for the unilateral ones. The results using the GMM estimator, which also combines both sources of identification, are very similar (column 4).

The results so far used the qualified sample, that is, the sample of individuals who know how to run a regression. Table~\ref{tab:alternative_samples} shows our main estimates for two alternative samples. Column 2 shows the results for the sample of all individuals. As expected, the proportion of control individuals spotting the error in the first question is smaller (6.8\% compared to 7.8\% in the qualified sample). The point estimate is also smaller (0.97pp versus 1.41pp). That is also expected; we cannot identify whether these new individuals correctly spot the 99 or not, because they likely do not know how to run OLS regressions, as they failed or did not answer question number 2. Indeed, we find that only 27.6\% of the unqualified sample provided a correct OLS point estimator (whether taking the 99 values into account or not), compared to 86.8\% for the qualified sample. In the third column, we drop from the qualified sample individuals who report negative values for the prior effect, as pre-specified in the PAP. The results look very similar, as only 3 individuals in the sample reported negative values. Lastly, in the fourth column, we restrict the qualified sample to the individuals who had priors close to the presented studies. Here, we consider those who reported priors in the interval 0.08--0.16 standard deviations. The point estimate is slightly larger, 1.54pp versus 1.41pp in the baseline estimation. We would expect this number to be larger as these individuals have priors aligned with the literature, and were, therefore, expecting positive results with the same magnitude that we present them.

\begin{table}[!ht]
    \centering
    \begin{threeparttable}
        \caption{Alternative Samples}\label{tab:alternative_samples}
    \label{tab:my_label}
    \begin{tabular}{lC{2.4cm}C{2.4cm}C{2.4cm}C{2.4cm}}
    \hline \hline 
    \addlinespace
Sample & Qualified & All & Non-negative Prior & Correct Prior \\ 
 & (1) & (2) & (3) & (4)   \\ 
\addlinespace
\hline 
\addlinespace
Effect           & \num{0.0141}       & \num{0.0097}       & \num{0.0141}       & \num{0.0154}       \\
(s.e.)& (\num{0.0073})     & (\num{0.0065})     & (\num{0.0073})     & (\num{0.0082})     \\
\addlinespace

[p-value]& [\num{0.0514}]     & [\num{0.1237}]     & [\num{0.0498}]     & [\num{0.0541}]     \\
\{p-value2\}& \{\num{0.0278}\} & \{\num{0.0615}\} & \{\num{0.0289}\} & \{\num{0.0299}\} \\

\\
N Obs          & \num{788}          & \num{944}          & \num{785}          & \num{697}                 \\
\\
Spotted First  & \num{ 0.078}       & \num{ 0.068}       & \num{ 0.078}       & \num{ 0.085}          \\
Spotted Second & \num{ 0.005}       & \num{ 0.007}       & \num{ 0.005}       & \num{ 0.006}      \\

\addlinespace
\hline \hline 
\end{tabular}
    \begin{tablenotes}
        \item \footnotesize{Notes: Table presents the results for the combined estimator for four different samples. The first column presents the results for the benchmark sample, defined in the main text as the qualified sample --- with all test takers who know how to run an OLS regression. The second column uses the entire sample. The third sample drops individuals with negative priors and the fourth column uses the sample of individuals who had prior beliefs between 0.08 and 0.16 standard deviations. The numbers in parentheses are the robust standard errors. In brackets and curly brackets, the p-values for the bilateral and unilateral tests were obtained using bootstrap with 30,000 replications.}
    \end{tablenotes}
        \end{threeparttable}
\end{table}

\subsection{Heterogeneity}

In this section, we investigate whether we have evidence of heterogeneous effects along some dimensions we observe in the data. For this exercise we use the entire sample for two main reasons. First to have larger groups when we split the analysis in sub-samples. Second, some exercises aim to compare individuals with lower or higher coding ability and skills, and therefore it would be inconsistent to already select on those that scored some screening questions correctly, as we do in the qualified sample. Nevertheless, Table~\ref{tab:heterogeneities_app} show the same results for the qualified sample. The first two rows of Table~\ref{tab:heterogeneities} show the benchmark results for the entire and qualified samples. As the subsequent panels use the entire sample, the first row should be used as a comparison.

\begin{table}[!ht]
    \centering
    \begin{threeparttable}
        \caption{Heterogeneities}\label{tab:heterogeneities}
    \label{tab:my_label}
    \begin{tabular}{L{3.3cm}C{1.5cm}C{1.5cm}C{1.5cm}C{1.5cm}C{1.5cm}C{1.5cm}}
    \hline \hline 
    \addlinespace
 & Effect & Std Error & P-value & N Obs & Spotted First & Diff P-value \\ 
\addlinespace
\hline 
\addlinespace
\multicolumn{6}{l}{\bfseries Benchmarks} \\ 
 \hphantom{a}Entire Sample & 0.0097 & (0.0065) & [0.1237] & 944 & 0.068 & - \\  
 \hphantom{a}Qualified Sample & 0.0141 & (0.0073) & [0.0514] & 788 & 0.078 & - \\ 
\\
\multicolumn{6}{l}{\bfseries 1. Clustered SE?} \\ 
\hphantom{a}No & -0.0079&(0.0109)&[0.3510]&155&0.014&\multirow{2}{*}{0.0981}\\
\hphantom{a}Yes & 0.0138&(0.0073)&[0.0552]&789&0.077&\\
\\ 
\multicolumn{6}{l}{\bfseries 2. Score in Q1--Q2} \\ 
\hphantom{a}Below Median & 0.0060&(0.0088)&[0.4852]&452&0.039&\multirow{2}{*}{0.5384}\\
\hphantom{a}Above Median &0.0137&(0.0089)&[0.0987]&489&0.096&\\
\\
\multicolumn{6}{l}{\bfseries 3. Coding Score} \\ 
\hphantom{a}Below Median  & -0.0021&(0.0100)&[0.8312]&405&0.062&\multirow{2}{*}{0.3205}\\
\hphantom{a}Above Median &0.0112&(0.0089)&[0.2089]&432&0.097&\\
\\
\multicolumn{6}{l}{\bfseries 4. Impact Evaluation Score} \\ 
\hphantom{a}Below Median  & 0.0043&(0.0107)&[0.6709]&423&0.087&\multirow{2}{*}{0.6365}\\
\hphantom{a}Above Median &0.0107&(0.0083)&[0.1982]&454&0.063&\\
\\
\multicolumn{6}{l}{\bfseries 5. Master's degree} \\ 
\hphantom{a}Yes & 0.0109&(0.0069)&[0.1094]&888&0.069&-\\
\\
\multicolumn{6}{l}{\bfseries 6. Econometrics course} \\ 
\hphantom{a}Yes &0.0111&(0.0074)&[0.1290]&830&0.076&-\\
\\
\multicolumn{6}{l}{\bfseries 7. Gender} \\ 
\hphantom{a}Men  &0.0083&(0.0071)&[0.1267]&578&0.074&\multirow{2}{*}{0.6713}\\
\hphantom{a}Women &0.0144&(0.0125)&[0.2453]&356&0.058&\\

\addlinespace
\hline \hline 
\end{tabular}
    \begin{tablenotes}
        \item \footnotesize{Notes: Table presents the results for the combined estimator for different samples. The columns show respectively the point estimate, standard errors (using bootstrap with 30,000 replications) , p-value, number of observations, proportion of individuals spotting the error in the first question (in the control group), and the p-value testing whether the difference of estimates is different than zero. Each row represents the estimates for a different sub-sample. The first two lines are the benchmark results for the entire and qualified samples. Each panel investigates heterogeneity by whether individuals clustered the standard errors in question 2, score in the first two initial questions, coding score, score on the impact evaluation assessment, whether they have a master's degree, have taken an econometrics course, and by gender. In panels 5 and 6 we do not display results for those without masters or without econometrics because the sample is very small.}
    \end{tablenotes}
        \end{threeparttable}
\end{table}

The first three panels split the individuals according to different ways of assessing their coding abilities. Whether they have correctly clustered the standard errors in question 2 (panel 1), their score in the two initial screening questions (panel 2), and their coding score in the partner institute assessment (panel 3). Across the three analysis, the results are very similar. Groups we expect to have better performance (who know how to cluster standard errors and with higher initial scores) have higher probabilities of correctly taking the 99 missing values regardless of the results the coding error would generate. However, these advantages do not make them less vulnerable to the bias. On the contrary, the point estimates are larger for the more trained subgroups although the differences are only significant at 10\% for the comparison between those who did or did not correctly clustered the standard errors. The next three panels analyze the results by training and knowledge of econometrics and impact evaluation. We see similar results, those with more training and knowledge (higher impact evaluation scores, with master's degree, and who have already taken an econometrics course) have larger point estimates, although the differences are  not statistically different. In panels 5 and 6, we do not estimate the results for those without master's degree or who did not took an econometrics course because the sample is so small that we dot enough variation to estimate the combined estimator. 

The last panel show heterogeneity by gender, where we observe that women have slightly lower baseline detection probability and larger point estimates for the bias. However, we cannot reject that the results are the same for both samples. In addition to these results, Appendix Table \ref{tab:waves} shows results separately for each wave.

\section{Discussion}

Our main result shows that individuals are significantly more likely to detect coding errors when those errors lead to unexpected results. This suggests that error detection is not a neutral process: it depends on whether the output aligns with prior expectations. In our setting, where the same coding error leads to either an expected or unexpected result depending on random assignment, we find that the unexpected result prompts greater debugging effort.

While our experimental design focuses on whether coding errors lead to expected or unexpected results, a natural conjecture is that this mechanism may also extend to \textit{favorable} results---that is, results that researchers view as more likely to be published or that support their hypotheses. If researchers are less inclined to scrutinize favorable outcomes, then coding errors that generate such results may be less likely to be detected, potentially introducing systematic bias into the published literature.

Our findings are particularly relevant for placebo tests, where researchers typically expect to find no significant effects---and where expected results are often also seen as favorable. In such cases, if a coding error leads to a statistically insignificant placebo result, researchers may interpret this as confirmation that the test ``worked'' and may forgo further scrutiny. As a consequence, coding errors may lead to an excess of false-negative findings, masking potential violations of identifying assumptions even when researchers act in good faith. 

More broadly, our results indicate that debugging is a costly activity, and thus the amount of effort researchers invest in it may depend on both their expectations and their incentives. Institutional practices---such as requiring code disclosure or pre-publication code review---may therefore significantly influence researchers' debugging efforts by altering their incentives. By increasing anticipated scrutiny, such practices could encourage more thorough error detection, potentially reducing biases arising from undetected mistakes.

\section{Conclusion}

In recent years, the Economics profession has seen increased concerns about the reproducibility and replicability of research findings. Many journals set up policies requiring data and code availability to increase research transparency. In this paper, we experimentally test whether individuals are more likely to find coding errors when they lead to non-expected results. We find that the probability of spotting a simple and common coding error increases by almost 20\% when the error leads to an unexpected result. This indicates that coding errors may not only increase the 
dispersion of results observed in empirical research but may bias the scientific inquiry. The results reinforce the necessity of policies that increase transparency in empirical science.

\singlespacing
\bibliography{codingbias_bib.bib}


\clearpage
\appendix
\section*{Online Appendices}

\doublespacing 
\addcontentsline{toc}{section}{Appendices}
\titleformat{\subsection}{\normalfont\fontsize{12}{12} \bfseries \center }{Appendix \Alph{subsection} - }{0em}{}
\renewcommand{\thesubsection}{\Alph{subsection}}
\titlespacing{\subsubsection}{0pt}{0cm}{0cm}

\setcounter{table}{0}
\renewcommand\thetable{A.\arabic{table}}

\setcounter{figure}{0}
\renewcommand\thefigure{A.\arabic{figure}}

\section{Additional figures and tables}

\begin{figure}[!ht]
    \centering
    \includegraphics[width=\linewidth]{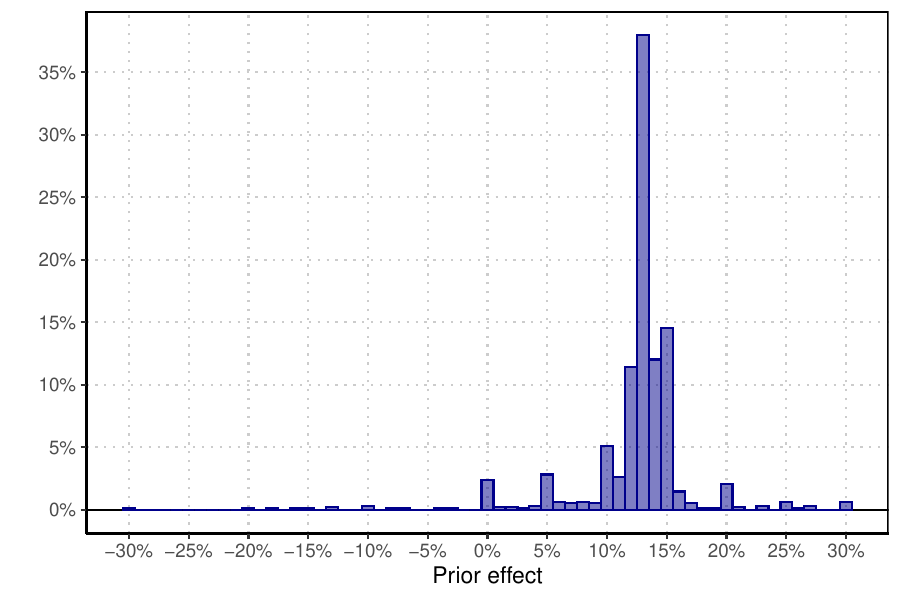}
    \caption{Prior effect}
    \label{fig:prior}
    \caption*{\footnotesize{Notes: The figure shows a histogram of the prior effect reported by the task takers. Each bar corresponds to the proportion of individuals reporting a prior at a given value, with a bandwidth of 1\%.}}
\end{figure}

\begin{table}[!ht]
    \centering
    \begin{threeparttable}
        \caption{Descriptive statistics and balance --- Qualified sample}\label{tab:balance_qualified}
    \label{tab:my_label}
    \begin{tabular}{lC{1.5cm}C{1.5cm}C{1.5cm}C{1.5cm}C{1.5cm}C{1.5cm}}
    \hline \hline 
    \addlinespace
Variable & Overall Mean & Control Mean & Treated Mean & Diff & p-value Diff & N Obs\\
\addlinespace
\hline 
\addlinespace
Female & 0.389 & 0.366 & 0.413 & 0.047 & 0.176 & 800\\
 & (0.488) & (0.482) & (0.493) & {}[0.035] &  & \\
 \addlinespace
Master & 0.953 & 0.933 & 0.974 & 0.041 & 0.005 & 806\\
 & (0.212) & (0.250) & (0.159) & {}[0.015] &  & \\
 \addlinespace
Econometrics & 0.907 & 0.897 & 0.918 & 0.020 & 0.318 & 806\\
 & (0.291) & (0.304) & (0.275) & {}[0.020] &  & \\
 \addlinespace
Stata & 0.537 & 0.542 & 0.531 & -0.011 & 0.758 & 807\\
 & (0.499) & (0.499) & (0.500) & {}[0.035] &  & \\
 \addlinespace
R & 0.307 & 0.310 & 0.304 & -0.006 & 0.850 & 807\\
 & (0.462) & (0.463) & (0.461) & {}[0.033] &  & \\
\addlinespace
Python & 0.143 & 0.131 & 0.155 & 0.023 & 0.344 & 807\\
 & (0.350) & (0.338) & (0.362) & {}[0.025] &  & \\
 \addlinespace
 Score Q1-Q2 & 5.133 & 5.121 & 5.145 & 0.024 & 0.744 & 807\\
 & (1.034) & (1.019) & (1.050) & {}[0.073] &  & \\
 \addlinespace
Coding Score & 0.150 & 0.100 & 0.201 & 0.101 & 0.179 & 733\\
 & (1.019) & (0.979) & (1.057) & {}[0.075] &  & \\
 \addlinespace
Impact Evaluation Score & 0.158 & 0.140 & 0.176 & 0.036 & 0.610 & 751\\
 & (0.963) & (0.936) & (0.992) & {}[0.070] &  & \\
 \addlinespace
Prior effect & 0.130 & 0.131 & 0.128 & -0.003 & 0.229 & 767\\
 & (0.034) & (0.030) & (0.037) & {}[0.002] &  & \\
\\
P-value joint test & & & & & 0.606 & \\
\addlinespace
\hline \hline 
\end{tabular}
    \begin{tablenotes}
        \item \footnotesize{Notes: The table shows the overall average, the average for the control group, the average for the treated group, the difference between the treated and control groups, the regression-adjusted p-value testing the equality of means for the treated and control groups, and the number of observations with non-missing values for each variable. The number in parenthesis is the standard deviation for each variable. The number in brackets shows the standard error of the regression-adjusted difference. The last row shows the p-value for the joint test that means of all these variables are the same between treated and control groups restricted to the qualified sample. Number of observations for each variable varies due to non-response to the specific question.}
    \end{tablenotes}
        \end{threeparttable}
\end{table}

\begin{table}[!ht]
    \centering
    \begin{threeparttable}
        \caption{Non-response rates}\label{tab:non_response}
    \label{tab:my_label}
    \begin{tabular}{lC{1.5cm}C{1.5cm}C{1.5cm}C{1.5cm}C{1.5cm}C{1.5cm}}
    \hline \hline 
    \addlinespace
Variable  & Control Mean & Treated Mean & Diff & P-value Diff & N Obs\\
\addlinespace
\hline 
\addlinespace
Female & 0.008 & 0.014 & 0.006 & 0.336 & 1036\\
 &  &  & {}[0.006] &  & \\
 \addlinespace
Master & 0.006 & 0.004 & -0.002 & 0.682 & 1036\\
 &  &  & {}[0.004] &  & \\
 \addlinespace
Econometrics & 0.004 & 0.000 & -0.004 & 0.157 & 1036\\
 &  &  & {}[0.003] &  & \\
 \addlinespace
Coding Score & 0.161 & 0.128 & -0.034 & 0.124 & 1036\\
 &  &  & {}[0.022] &  & \\
 \addlinespace
Impact Evaluation Score & 0.095 & 0.073 & -0.022 & 0.197 & 1036\\
 &  &  & {}[0.017] &  & \\
\addlinespace
Prior effect & 0.085 & 0.086 & 0.001 & 0.952 & 1036\\
 &  &  & {}[0.017] &  & \\
\addlinespace
\hline \hline 
\end{tabular}
    \begin{tablenotes}
        \item \footnotesize{Notes: The table shows the non-response rates for the control and treatment groups, the differential rate between the treated and control groups, the regression-adjusted p-value testing the equality of means for the treated and control groups, and the number of observations for each variable. The number in brackets shows the standard error of the regression-adjusted difference.}
    \end{tablenotes}
        \end{threeparttable}
\end{table}

\begin{table}[!ht]
    \centering
    \begin{threeparttable}
        \caption{Results separately by wave}\label{tab:waves}
    \label{tab:my_label}
    \begin{tabular}{lC{2.2cm}C{2.2cm}C{2.2cm}}
    \hline \hline 
    \addlinespace
Sample & All & Wave 2024 & Wave 2025 \\ 
 & (1) & (2) \\ 
\addlinespace
\hline 
\addlinespace
Effect           & \num{0.0141}       & \num{0.0172}       & \num{0.0093}       \\
(s.e.)& (\num{0.0073})     & (\num{0.0077})     & (\num{0.0135})     \\
\addlinespace

[p-value]& [\num{0.0514}]     & [\num{0.0273}]     & [\num{0.4652}]     \\
\{p-value2\}& \{\num{0.0278}\} & \{\num{0.0224}\} & \{\num{0.2285}\} \\
\\
N Obs          & \num{788}          & \num{450}          & \num{338}          \\
\addlinespace
\hline \hline 
\end{tabular}
    \begin{tablenotes}
        \item \footnotesize{Notes: Table presents the results for the combined estimator for two different samples. The first column presents the results for the 2024 wave sample and the second column for the 2025 wave. The numbers in parentheses are the robust standard errors. In brackets and curly brackets, the p-values for the bilateral and unilateral tests were obtained using bootstrap with 30,000 replications.}
    \end{tablenotes}
        \end{threeparttable}
\end{table}

\begin{table}[!ht]
    \centering
    \begin{threeparttable}
        \caption{Heterogeneities --- qualified sample}\label{tab:heterogeneities_app}
    \label{tab:my_label}
    \begin{tabular}{L{3.3cm}C{1.5cm}C{1.5cm}C{1.5cm}C{1.5cm}C{1.5cm}C{1.5cm}}
    \hline \hline 
    \addlinespace
 & Effect & Std Error & P-value & N Obs & Spotted First & Diff P-value \\ 
\addlinespace
\hline 
\addlinespace
\multicolumn{6}{l}{\bfseries Benchmarks} \\ 
 \hphantom{a}Entire Sample & 0.0097 & (0.0065) & [0.1237] & 944 & 0.068 & - \\  
 \hphantom{a}Qualified Sample & 0.0141 & (0.0073) & [0.0514] & 788 & 0.078 & - \\ 
\\
\multicolumn{6}{l}{\bfseries 1. Clustered SE?} \\ 
\hphantom{a}Yes & 0.0145&(0.0077)&[0.052]&747&0.081&\\
\\ 
\multicolumn{6}{l}{\bfseries 2. Score in Q1--Q2} \\ 
\hphantom{a}Below Median &0.0131&(0.0093)&[0.1763]&390&0.043&\multirow{2}{*}{0.9441}\\
\hphantom{a}Above Median &0.0141&(0.0108)&[0.1461]&398&0.113&\\
\\
\multicolumn{6}{l}{\bfseries 3. Coding Score} \\ 
\hphantom{a}Below Median  & 0.0035&(0.0100)&[0.7003]&362&0.074&\multirow{2}{*}{0.4690}\\
\hphantom{a}Above Median &0.014&(0.0105)&[0.2005]&362&0.102&\\
\\
\multicolumn{6}{l}{\bfseries 4. Impact Evaluation Score} \\ 
\hphantom{a}Below Median  &0.0154&(0.0121)&[0.1637]&365&0.087&\multirow{2}{*}{1.3871}\\
\hphantom{a}Above Median &0.0078&(0.0089)&[0.4067]&371&0.063&\\
\\
\multicolumn{6}{l}{\bfseries 5. Master's degree} \\ 
\hphantom{a}Yes &0.0154&(0.0077)&[0.0454]&750&0.078&-\\
\\
\multicolumn{6}{l}{\bfseries 6. Econometrics course} \\ 
\hphantom{a}Yes &0.0156&(0.0081)&[0.0476]&714&0.084&-\\
\\
\multicolumn{6}{l}{\bfseries 7. Gender} \\ 
\hphantom{a}Men  &0.0101&(0.0085)&[0.2128]&477&0.089&\multirow{2}{*}{0.4519}\\
\hphantom{a}Women &0.0221&(0.0135)&[0.1055]&304&0.061&\\

\addlinespace
\hline \hline 
\end{tabular}
    \begin{tablenotes}
        \item \footnotesize{Notes: Table presents the results for the combined estimator for different samples. The columns show respectively the point estimate, standard errors (using bootstrap with 30,000 replications) , p-value, number of observations, proportion of individuals spotting the error in the first question (in the control group), and the p-value testing whether the difference of estimates is different than zero. Each row represents the estimates for a different sub-sample. The first two lines are the benchmark results for the entire and qualified samples. Each panel investigates heterogeneity by whether individuals clustered the standard errors in question 2, score in the first two initial questions, coding score, score on the impact evaluation assessment, whether they have a master's degree, have taken an econometrics course, and by gender. In panels 1, 5, and 6 we do not display results for those who did not cluster, without masters or without econometrics because the sample is very small.}
    \end{tablenotes}
        \end{threeparttable}
\end{table}

\clearpage
\section{Latent types expanded}\label{app:latent}

In the main text we present our identification strategy, classifying individuals into four latent types. In this appendix we expand this classification in order to relax the hypothesis the Complier II type (CII) detects the error irrespectively of the order of results. For this we sub-divide this type into three exhaustive cases. In this case, we consider that there are six latent types:

\begin{enumerate}
    \item Always-spot (AS): those who always spot the error, irrespectively of the result;

    \item Never-spot (NS): those who never spot the error, irrespectively of the result; 

    \item Complier I (CI): those who spot the coding error if it leads to unexpected results;

    \item Complier II (CII): those who spot the coding error if they find conflicting results between the two answers 
    \begin{itemize}
        \item CII-A: spot the error if they find conflicting results between the two answers irrespectively of their signs
        \item CII-B: spot the error if they find conflicting results between the two answers, if the first is the positive and the second is the negative one. But not the other way around.
        \item CII-C: spot the error if they find conflicting results between the two answers, if the first is the negative and the second is the positive one. But not the other way around.
    \end{itemize}    
\end{enumerate}

\begin{figure}[!h]
\caption{Latent types and identification (expanded)}\label{fig:latent_types_expanded}
\begin{center}
\begin{minipage}{0.48\textwidth}
\centering
\begin{tikzpicture}
\draw (0,0) rectangle (6,4);
\draw (3,0) -- (3,4); 
\draw (0,2) -- (6,2); 

\node at (1.5, 4.3) {\footnotesize Spotted};
\node at (4.5, 4.3) {\footnotesize Not Spotted};
\node at (3, 4.8) {Q4};
\node at (3, 5.6) {\bfseries Control};

\node[rotate=90] at (-0.3, 3) {\footnotesize Spotted};
\node[rotate=90] at (-0.3, 1) {\footnotesize Not Spotted};
\node[rotate=90] at (-1, 2) {Q3};

\node at (1.5, 3) {\footnotesize AS};
\node at (1.5, 1) {\footnotesize CI, CII-A, CII-B};
\node at (4.5, 1) {\footnotesize NS, CII-C};
\node at (4.5, 3) {-};
\end{tikzpicture}
\end{minipage}
\hfill
\begin{minipage}{0.48\textwidth}
\centering
\begin{tikzpicture}
\draw (0,0) rectangle (6,4);
\draw (3,0) -- (3,4); 
\draw (0,2) -- (6,2); 

\node at (1.5, 4.3) {\footnotesize Spotted};
\node at (4.5, 4.3) {\footnotesize Not Spotted};
\node at (3, 4.8) {Q4};
\node at (3, 5.6) {\bfseries Treated};

\node[rotate=90] at (-0.3, 3) {\footnotesize Spotted};
\node[rotate=90] at (-0.3, 1) {\footnotesize Not Spotted};
\node[rotate=90] at (-1, 2) {Q3};

\node at (1.5, 3) {\footnotesize AS, CI};
\node at (1.5, 1) {\footnotesize CII-A, CII-C};
\node at (4.5, 1) {\footnotesize NS, CII-B};
\node at (4.5, 3) {-};
\end{tikzpicture}
\end{minipage}
\end{center}
\end{figure}

Figure~\ref{fig:latent_types_expanded} shows all the latent types by their respective results in Q3 and Q4 if they are in the treatment or control group. First, note that our first source of identification --- which compares the proportion of treated and controls who spotted the error in Q3 --- is not affected by the presence of the sub-categories of CII types.

For the second identification approach --- which contrasts those spotting only in Q4 between treatment and control, we  would identify the following quantity $\Delta_{2}$:
\begin{equation}
\begin{aligned}
\Delta_2 &= (\text{CI}+\text{CII-A}+\text{CII-B})-(\text{CII-A}-\text{CII-C}))\\
&= \text{CI}+(\text{CII-B}-\text{CII-C}).
\end{aligned}
\end{equation}
That is, we would identify the proportion of CI types, plus the difference between CII-B and CII-C types in the population. Therefore, this approach recovers the proportion of CI if these two types have the same proportion (or do not exist). If $\text{CII-B}<\text{CII-C}$, that is, it is more likely to spot the error after seeing negative-positive, than positive-negative results, then we would underestimate our target parameter. If $\text{CII-B}>\text{CII-C}$, we would overestimate the proportion of CI types. Note however, that this difference is also manifested by individuals with differential debugging probabilities based on the results they face, exactly what we want to test with this RCT. Therefore, we do not see that as necessarily a bias, but as an evidence that the debugging probabilities depend on the observed outcome when there is a coding error. Additionally, it is worth mentioning that the proportion of the sum of CII-A and CII-C types is identified by the proportion of individuals spotting only in Q4 in the treatment group. Since this proportion is very small (0.5\%), this implies that the proportion of CII-C is also very small.

\section{Data Task}
\label{Appendix_datatask}

Figures~\ref{dt_1}---\ref{dt_12} below reproduce the six parts of the data task as seen by the task takers.

\bigskip 

\begin{figure}[ht!]
    \centering
\includegraphics[width=\textwidth]{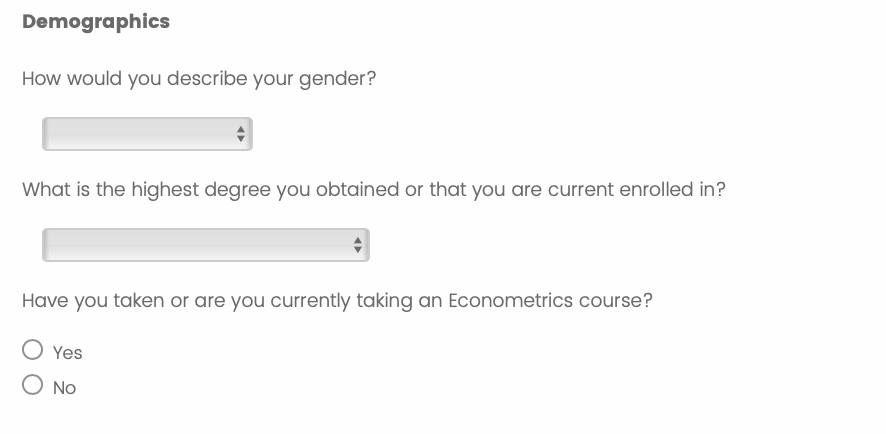}
    \caption{Part I - Demographics}
    \label{dt_1}
\end{figure}

\begin{figure}[ht!]
    \centering
\includegraphics[width=\textwidth]{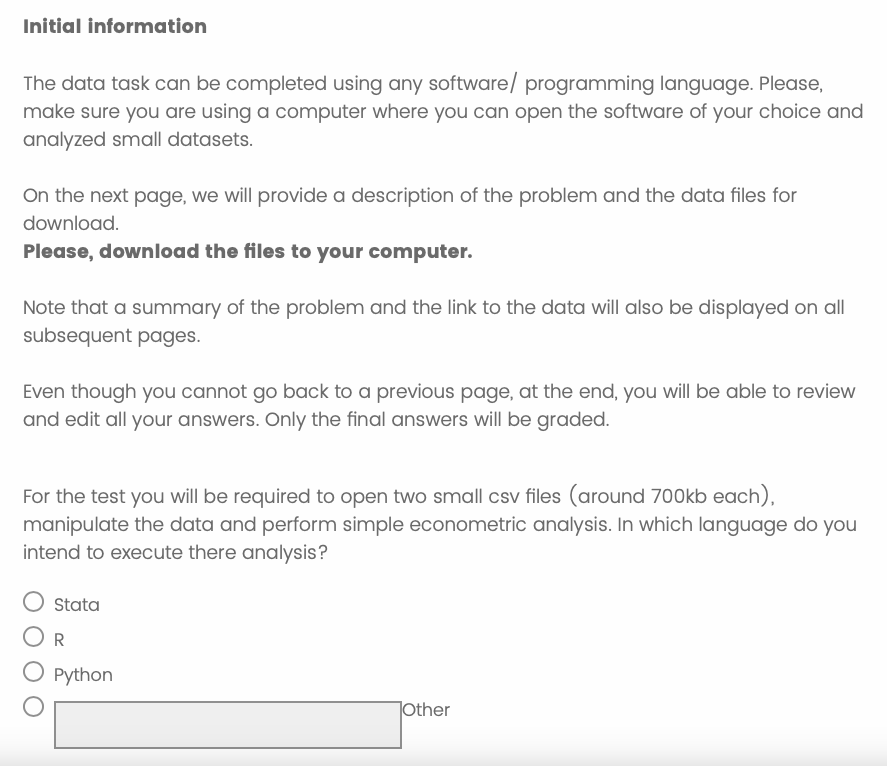}
    \caption{Part I - Computational Language}
\end{figure}

\begin{figure}[ht!]
    \centering
\includegraphics[width=\textwidth]{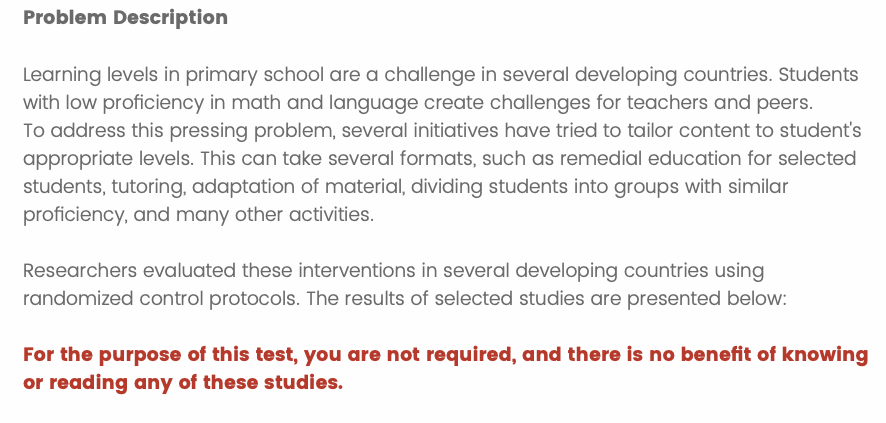}
    \caption{Part II - Literature (I)}
\end{figure}

\begin{figure}[ht!]
    \centering
\includegraphics[width=\textwidth]{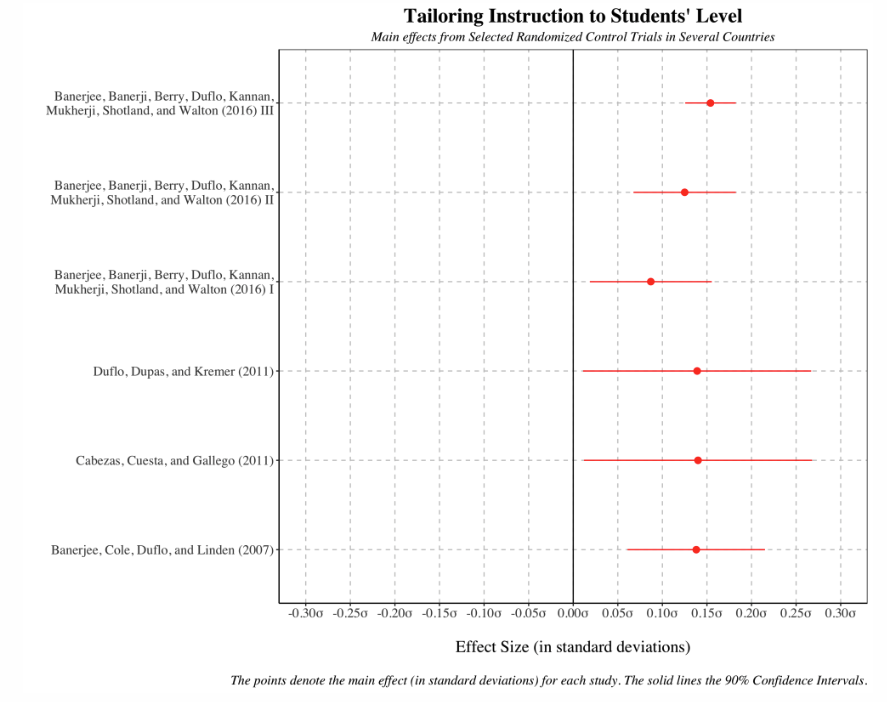}
    \caption{Part II - Literature (II)}
\end{figure}

\begin{figure}[ht!]
    \centering
\includegraphics[width=\textwidth]{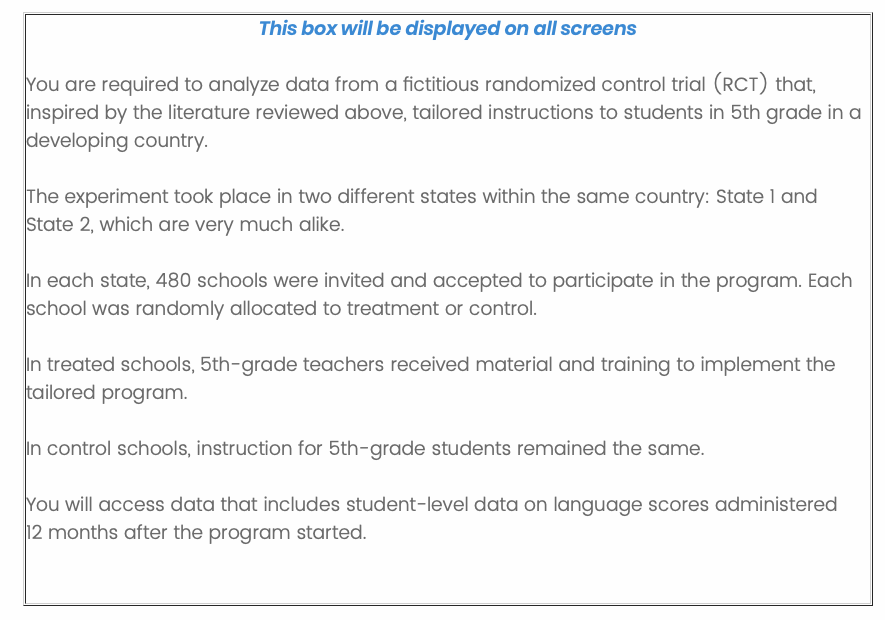}
    \caption{Part III - The example}
\end{figure}

\begin{figure}[ht!]
    \centering
\includegraphics[width=\textwidth]{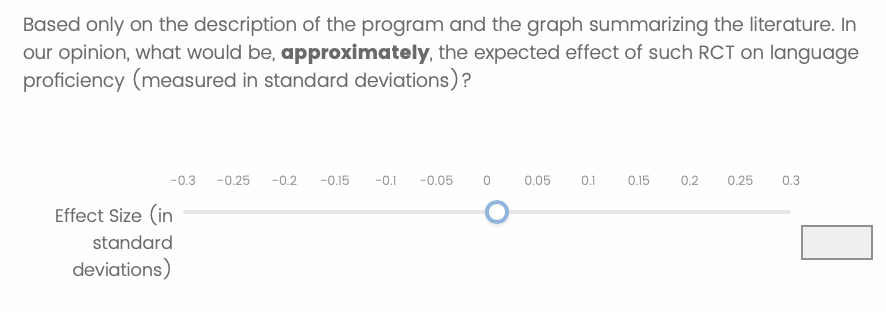}
    \caption{Part III - Eliciting priors}
\end{figure}

\begin{figure}[ht!]
    \centering
\includegraphics[width=\textwidth]{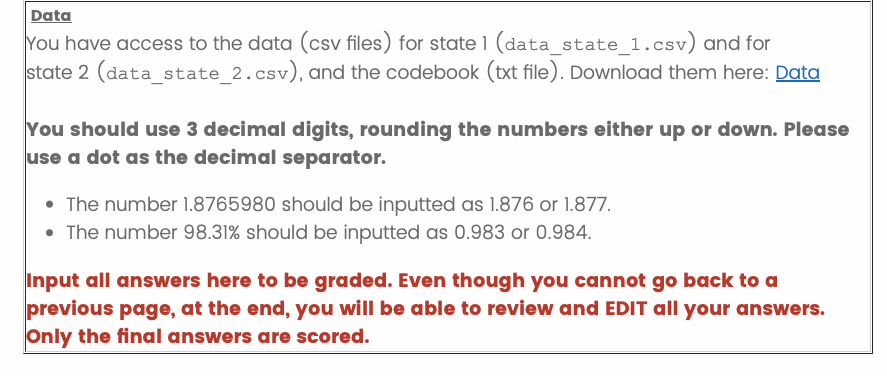}
    \caption{Part IV - The data}
\end{figure}

\begin{figure}[ht!]
    \centering
\includegraphics[width=\textwidth]{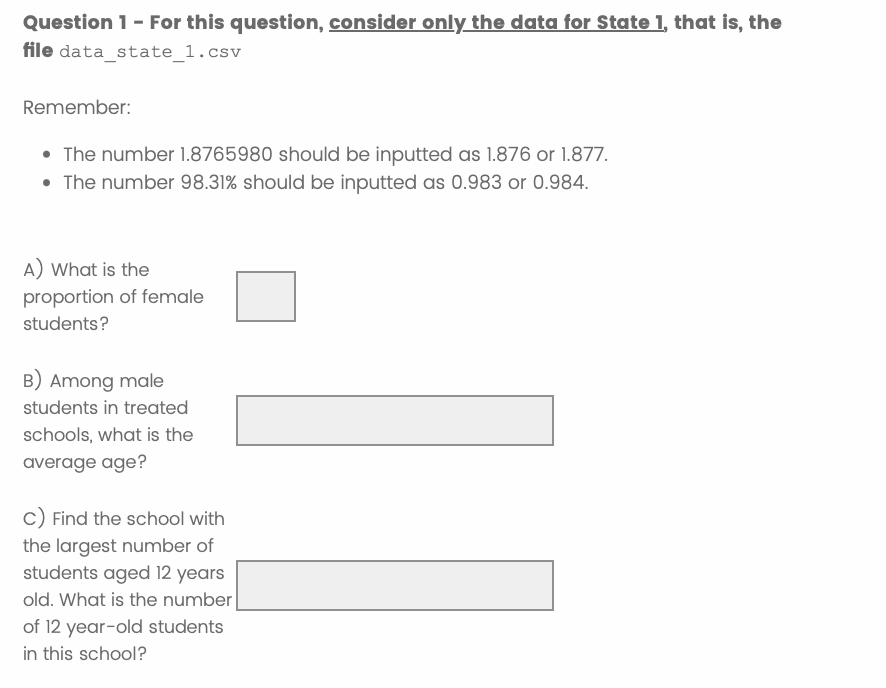}
    \caption{Part IV - Question 1}
\end{figure}

\begin{figure}[ht!]
    \centering
\includegraphics[width=\textwidth]{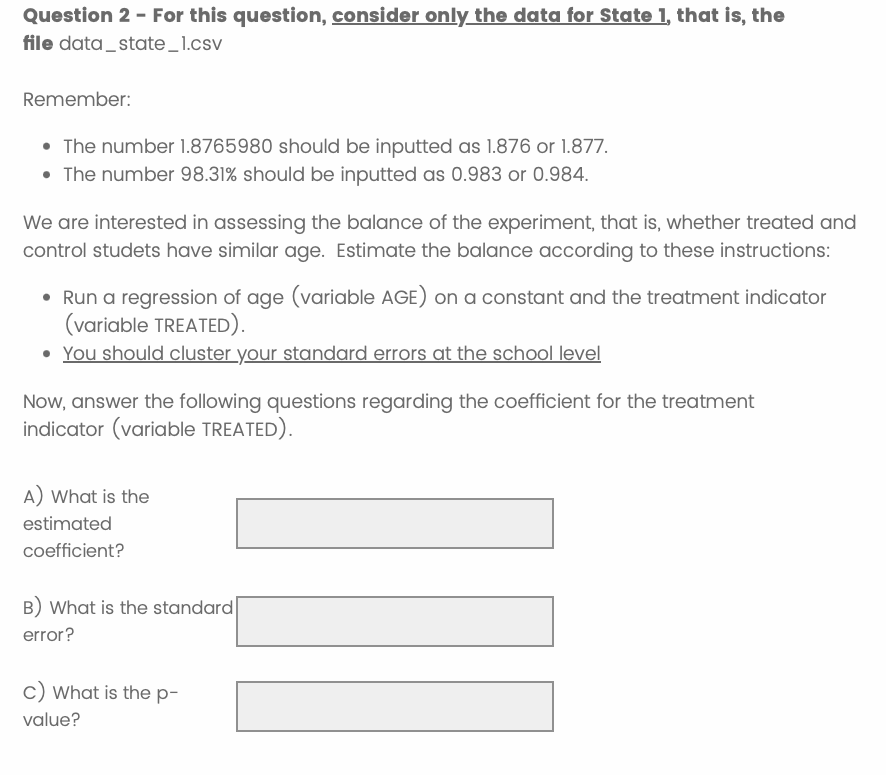}
    \caption{Part IV - Question 2}
\end{figure}

\begin{figure}[ht!]
    \centering
\includegraphics[width=\textwidth]{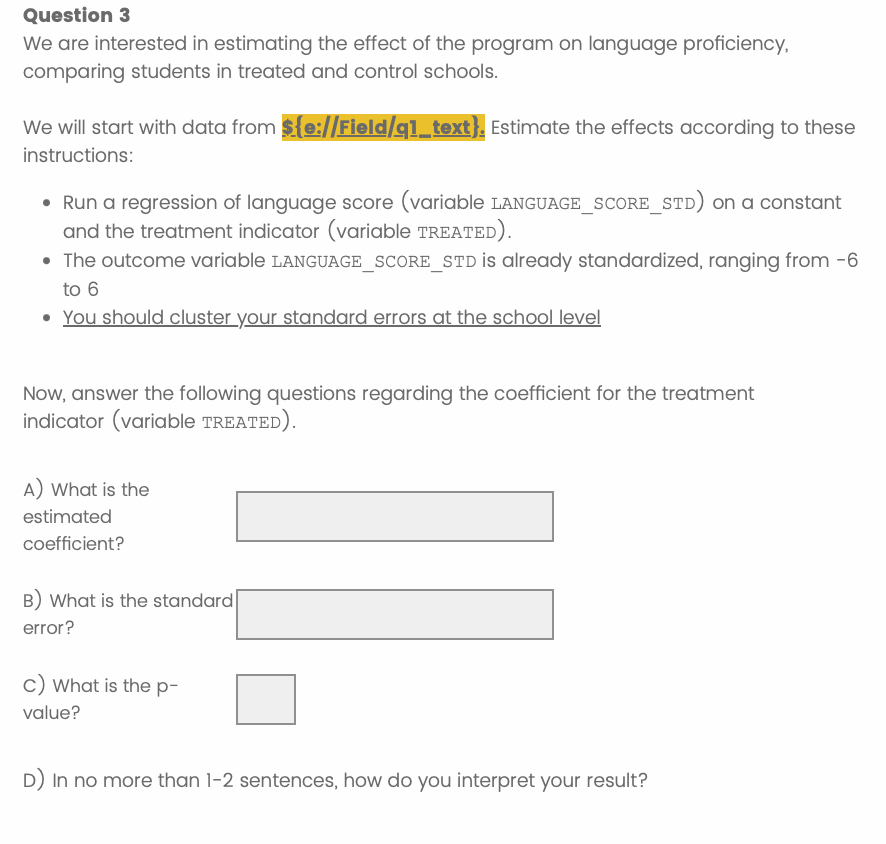}
    \caption{Part V - Question 3}
\end{figure}

\begin{figure}[ht!]
    \centering
\includegraphics[width=\textwidth]{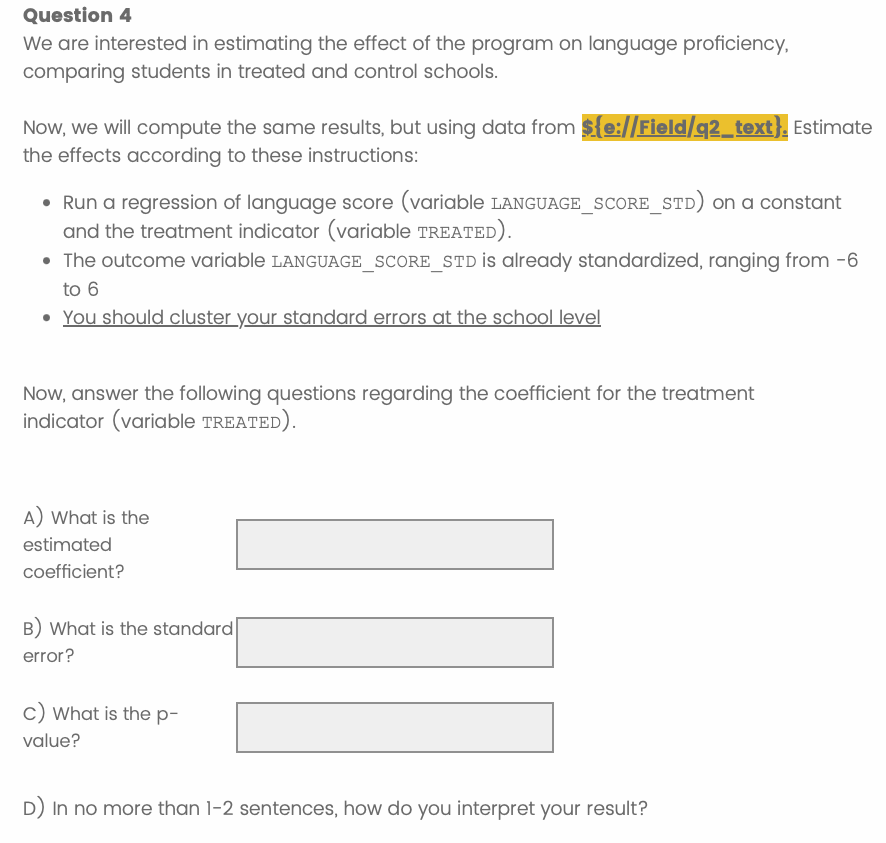}
    \caption{Part V - Question 4}
\end{figure}

\begin{figure}[ht!]
    \centering
\includegraphics[width=\textwidth]{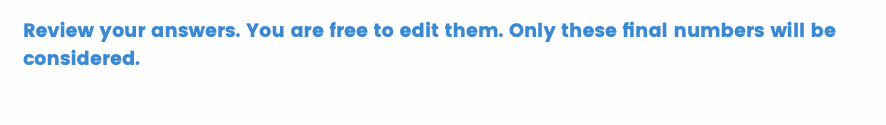}
    \caption{Part VI - Revising answers}
    \label{dt_12}
\end{figure}

\end{document}